\documentclass{ws-ijmpe}

\newcommand{\pT}{p_\mathrm{T}}
\newcommand{\kt}{k_\mathrm{T}}
\newcommand{\et}{E_\mathrm{T}}
\newcommand{\gev}{\mathrm{GeV}}
\newcommand{\gevc}{\mathrm{GeV}/c}

\begin{document}

\markboth{Jan Kapit\'an}{Jet measurements in p+p and d+Au collisions with STAR at RHIC}

\catchline{}{}{}{}{}

\title{JET MEASUREMENTS IN P+P AND D+AU COLLISIONS WITH STAR AT RHIC}

\author{\footnotesize JAN KAPIT\'AN (for the STAR Collaboration)}

\address{Nuclear Physics Institute ASCR, Na Truhlarce 39/64\\
18086 Praha 8\\
Czech Republic\\
kapitan@rcf.rhic.bnl.gov}

\maketitle

\begin{history}
\received{(received date)}
\revised{(revised date)}
\end{history}

\begin{abstract}
Full jet reconstruction in heavy-ion collisions is a promising tool for the quantitative study of properties of the dense medium produced at RHIC. In addition to baseline measurements in p+p, results from d+Au collisions are needed to disentangle initial state nuclear effects from medium-induced $\kt$ broadening and jet quenching. We present mid-rapidity inclusive jet $\pT$ spectra and di-jet correlations ($\kt$) in 200~GeV p+p and d+Au collisions from the 2007-2008 RHIC run.
\end{abstract}

\section{Introduction}
\label{intro}
Jets are collimated sprays of particles produced in high energy particle collisions. They are remnants of hard-scattered partons, which are the fundamental objects of pQCD. At RHIC, jets can be used as a probe of the hot and dense matter created in heavy ion collisions~\cite{MP,EB}. To quantify the signals observed in heavy ion collisions in comparison to p+p collisions, it is necessary to measure the cold nuclear matter effects in systems such as d+Au. 

Cold nuclear matter effects can be described by partonic rescattering~\cite{vitev} and by modification of parton distribution functions~\cite{eps}. Measurements of single particle $R_\mathrm{dAu}$ have shown modification even at midrapidity~\cite{singleRdAu}. It is important to verify these observations using jets: they provide much higher kinematic reach as they are not prone to fragmentation biases and one can also avoid uncertainties coming from imprecise knowledge of fragmentation functions. An alternative approach to full jet reconstruction is a method using di-hadron correlations, discussed in detail in~\cite{Mriganka}.

\section{Jet reconstruction}
\label{jets}
This analysis is based on $\sqrt{s_\mathrm{NN}} = 200~\gev$ data from the STAR experiment, recorded during RHIC run 8 (2007-2008). The Beam Beam Counter detector, located in the Au nucleus fragmentation region, was used to select the 20\% highest multiplicity events in d+Au collisions. The Barrel Electromagnetic Calorimeter (BEMC) detector was used to measure the neutral component of jets, and the Time Projection Chamber (TPC) detector was used to measure the charged component of jets. In the case of a TPC track pointing to a BEMC tower, its momentum was subtracted from the tower energy to avoid double counting (electrons, MIP and possible hadronic showers in the BEMC). 

To minimize the effect of BEMC backgrounds and dead areas, the jet neutral energy fraction is required to be within $(0.1,0.9)$. An upper $\pT < 15~\gevc$ cut was applied to TPC tracks due to uncertainties in TPC tracking performance at high-$\pT$ in run 8 (under further investigation). The acceptance of TPC and BEMC together with experimental details (calibration, primary vertex position cuts) limit the jet fiducial acceptance to $|\eta|<0.55 \; (R=0.4), |\eta|<0.4 \; (R=0.5)$, where $R$ is the resolution parameter used in jet finding.

Recombination jet algorithms kt and anti-kt, part of the FastJet package~\cite{fj}, are used for jet reconstruction. To subtract d+Au underlying event background, a method based on active jet areas~\cite{bgsub} is applied event-by-event: $\pT^{Rec} = \pT^{Candidate} - \rho \cdot A$, with $\rho$ estimating the background density per event and $A$ being the jet active area. Due to the asymmetry of the colliding d+Au system, the background is asymmetric in $\eta$. This dependence was fit with a linear function in $\eta$ and included in the background subtraction procedure. 

Pythia 6.410 and GEANT detector simulations were used to correct for experimental effects. Jet reconstruction was run at MC hadron level (PyMC) and at detector level (PyGe). To study residual effects of the d+Au background (such as background fluctuations), a sample with added background (PyBg) was created by mixing Pythia events with 0-20\% highest multiplicity d+Au events (minimum bias online trigger). This mixing was done at detector level (reconstructed TPC tracks and BEMC towers).

\section{Nuclear $\kt$ broadening}

Comparing azimuthal correlations of jets in di-jet events in p+p and d+Au can provide information on nuclear $\kt$ broadening. To increase di-jet yield, BEMC high tower (HT) online trigger was employed (one tower with $\et > 4.3~\gev$) for both p+p and d+Au data (run 8). Resolution parameter $R=0.5$ was used for jet finding and a cut $\pT > 0.5~\gevc$ applied for tracks and towers to reduce background. 

To select a clean di-jet sample two highest energy jets ($p_\mathrm{T,1} > p_\mathrm{T,2}$) in each event were used, with $p_\mathrm{T,2} > 10~\gevc$~\cite{qm09}. Distributions of $k_\mathrm{T,raw} = p_\mathrm{T,1} \sin(\Delta\phi)$ were constructed for di-jets and Gaussian widths, $\sigma_{k_\mathrm{T,raw}}$, were obtained for the two jet algorithms and two ($10 - 20~\gevc$, $20 - 30~\gevc$) $p_\mathrm{T,2}$ bins. 

\begin{figure}[htb]
\centering
\includegraphics[width=0.9\textwidth]{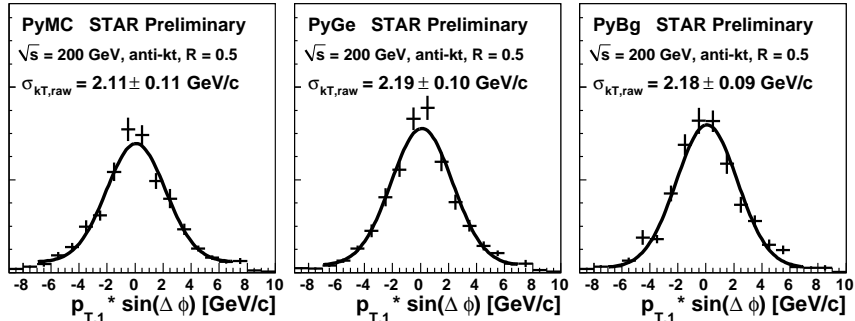}
\vspace{-0.5cm}
\caption{\label{fig:ktsimu}Distributions of $k_\mathrm{T,raw} = p_\mathrm{T,1} \sin(\Delta\phi)$ for simulation ($10 < p_\mathrm{T,2} < 20~\gevc$).}
\end{figure}

Detector and residual background effects on the $\kt$ widths were studied by comparing PyMC, PyGe and PyBg distributions as shown in Fig.~\ref{fig:ktsimu}. The widths are the same within statistical errors, most likely due to the interplay between jet $\pT$ and $\Delta\phi$ resolutions. Fig.~\ref{fig:ktdata} shows an example of the $k_\mathrm{T,raw}$ distributions for data. 
The Gaussian fit to p+p data is not ideal and the precise shape of the distribution is under study. RMS widths of these distributions have therefore been checked and they agree with the sigma widths of the fits.
The values extracted from the Gaussian fits are $\sigma_{k_\mathrm{T,raw}}^{p+p} = 2.8 \pm 0.1~\mathrm{(stat)}~\gevc$ and $\sigma_{k_\mathrm{T,raw}}^{d+Au} = 3.0 \pm 0.1~\mathrm{(stat)}~\gevc$. Possible nuclear $\kt$ broadening therefore seems rather small. 

The systematic uncertainties on extracted $\kt$ widths come from Jet Energy Scale (JES) uncertainty (discussed in more detail in section~\ref{systematics}) and from the way they were extracted. The latter is estimated to be $0.2~\gevc$ and includes a weak dependence on the $|\Delta\phi - \pi|$ cut for back-to-back di-jet selection (varied between 0.5 and 1.0), differences in the $p_\mathrm{T,2}$ range and jet algorithm selections and the precision with which the detector and background effects were found to be negligible. 

\section{Inclusive jet spectra}
\label{spectra}
Run 8 d+Au data with a minimum bias online trigger were used for this study. 10M 0-20\% highest multiplicity events after event cuts were used for jet finding (anti-kt algorithm) with resolution parameter $R = 0.4$, $\pT > 0.2~\gevc$ cut was applied to tracks and towers. The jet $\pT$ spectrum is normalized per event and the high multiplicity of d+Au events also guarantees the trigger efficiency is independent of the $\pT$ of the hard scattering. Therefore, no correction related to trigger is applied to jet $\pT$ spectra.

\begin{figure}[htb]
\begin{minipage}[h]{0.62\textwidth}
\includegraphics[width=\textwidth]{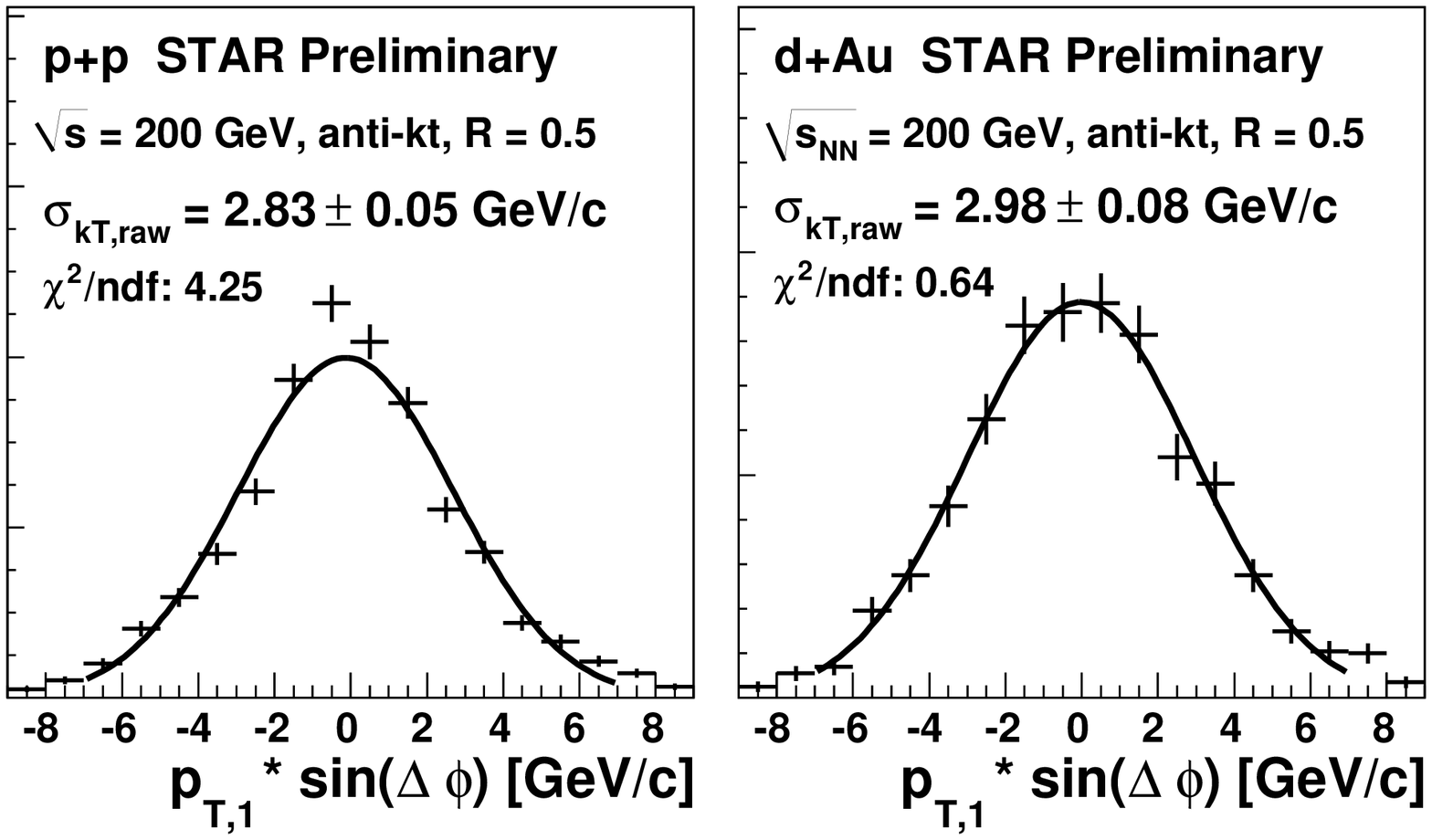}
\vspace{-0.95cm}
\caption{\label{fig:ktdata}Distributions of $k_\mathrm{T,raw}$ for p+p, d+Au ($10 < p_\mathrm{T,2} < 20~\gevc$).}
\end{minipage}
\hfill
\begin{minipage}[h]{0.33\textwidth}    
\includegraphics[width=\textwidth]{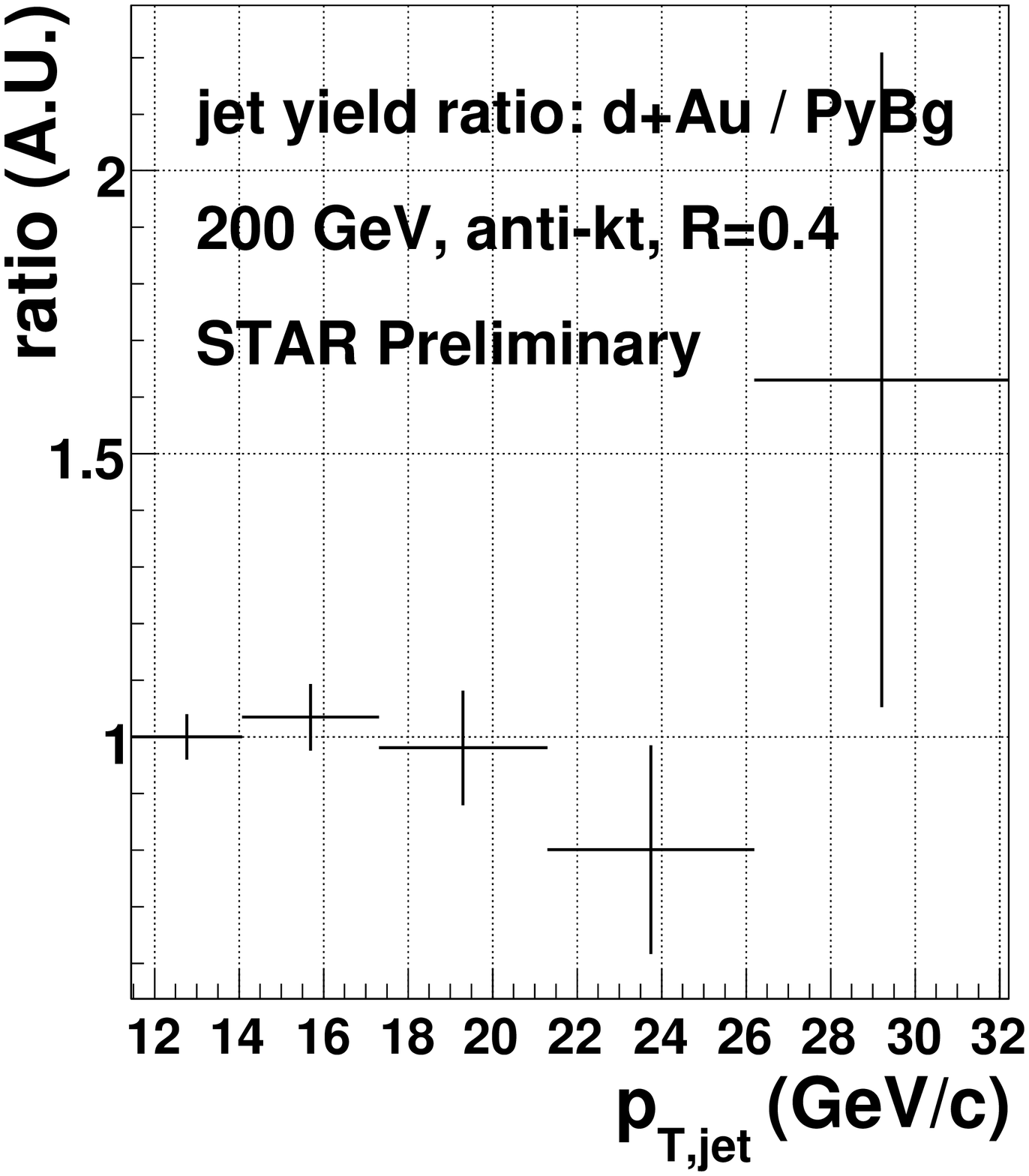}
\vspace{-0.99cm}
\caption{\label{fig:ratio}Ratio of jet $\pT$ spectra between d+Au and simulation.}
\end{minipage}
\end{figure}

A bin-by-bin correction is used to correct jet $\pT$ spectrum to hadron level. It is based on the generalized efficiency, constructed as the ratio of PyMC to PyBg jet $\pT$ spectra, applied to the measured jet $\pT$ spectrum. It therefore corrects for detector effects (tracking efficiency, unobserved neutral energy, jet $\pT$ resolution) as well as for residual background effects. As the impact of these effects on jet $\pT$ spectrum differ substantially depending on jet $\pT$ spectrum shape, the shapes have to be consistent between the PyBg and the measured jet $\pT$ spectra. Fig.~\ref{fig:ratio} shows that this is indeed the case.

As the jet $\pT$ spectrum is very sensitive to the jet energy scale, an additional correction was applied here to account for the lower TPC tracking efficiency in d+Au compared to that from the used p+p Pythia simulation. The d+Au efficiency was determined by simulating single pions and embedding them at the raw detector level into real d+Au minimum bias events. 
The tracking efficiency in the Pythia simulation was then artificially lowered, prior to jet finding at PyGe and PyBg level, so that it matches the one obtained from d+Au embedding. 

To compare per event jet yield in d+Au to jet cross section in p+p collisions, an input from MC Glauber study is utilized: $\langle N_\mathrm{bin} \rangle = 14.6 \pm 1.7$ for 0-20\% highest multiplicity d+Au collisions and $\sigma_\mathrm{inel,pp} = 42~\mathrm{mb}$. These factors were used to scale the p+p jet cross section measured previously by the STAR collaboration~\cite{ppjetprl} using Mid Point Cone (MPC) jet algorithm with $R = 0.4$.
The resulting d+Au jet $\pT$ spectrum is shown in Fig.~\ref{fig:spectrum} together with the scaled p+p jet spectrum. The systematic errors are indicated by dashed lines and by the gray boxes. The dominant contribution to p+p systematic uncertainty is the Jet Energy Scale (JES) uncertainty. Within these systematic uncertainties, the d+Au jet spectrum shows no significant deviation from the scaled p+p spectrum.

\section {Systematic uncertainties}
\label{systematics}
\begin{figure}[htb]
\begin{minipage}[h]{0.61\textwidth}
\centering
\includegraphics[width=\textwidth]{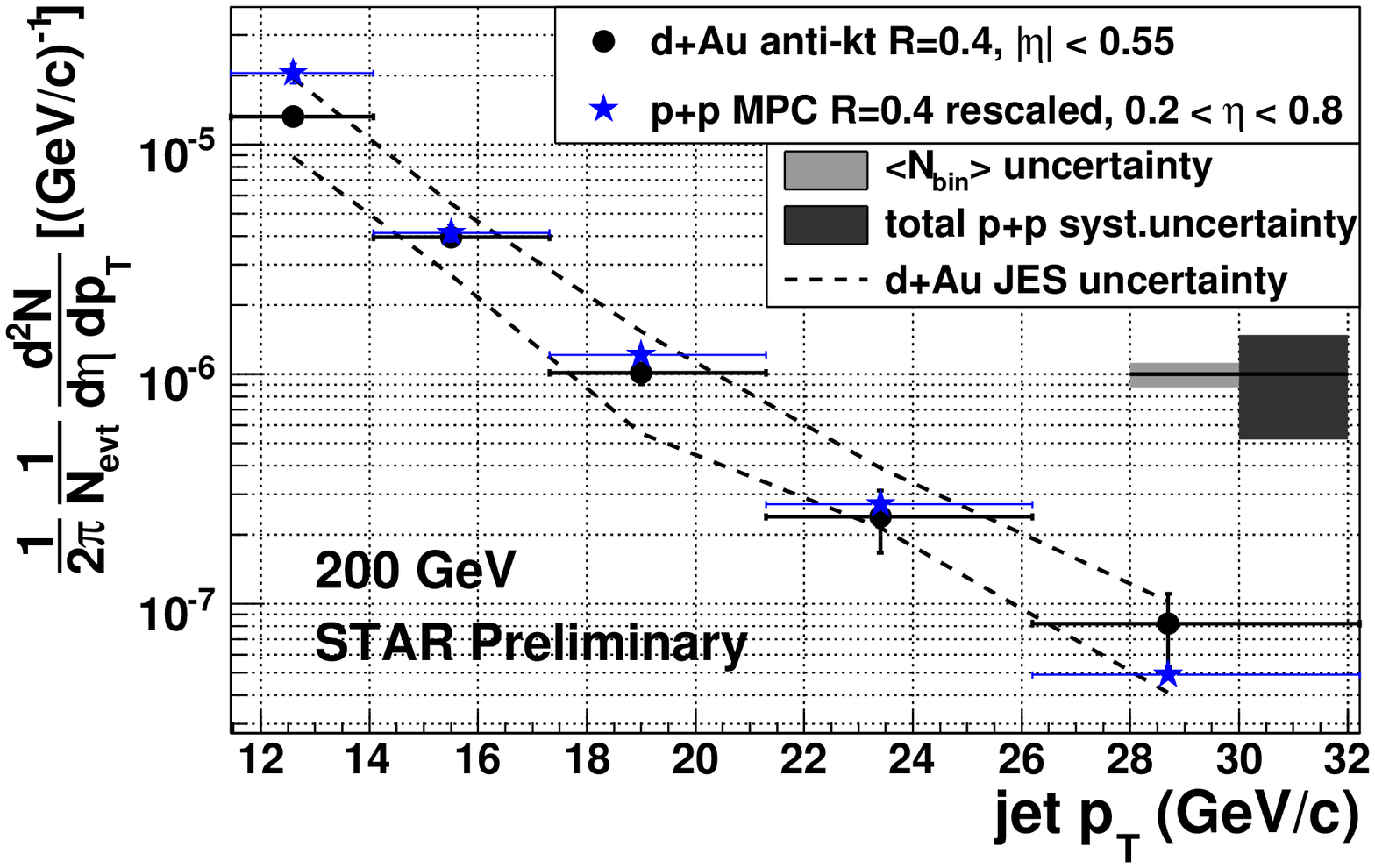}
\vspace{-0.9cm}
\caption{\label{fig:spectrum}Jet $\pT$ spectrum: d+Au collisions compared to scaled p+p~\protect\cite{ppjetprl}.}
\end{minipage}
\hfill
\begin{minipage}[h]{0.37\textwidth}
\includegraphics[width=\textwidth]{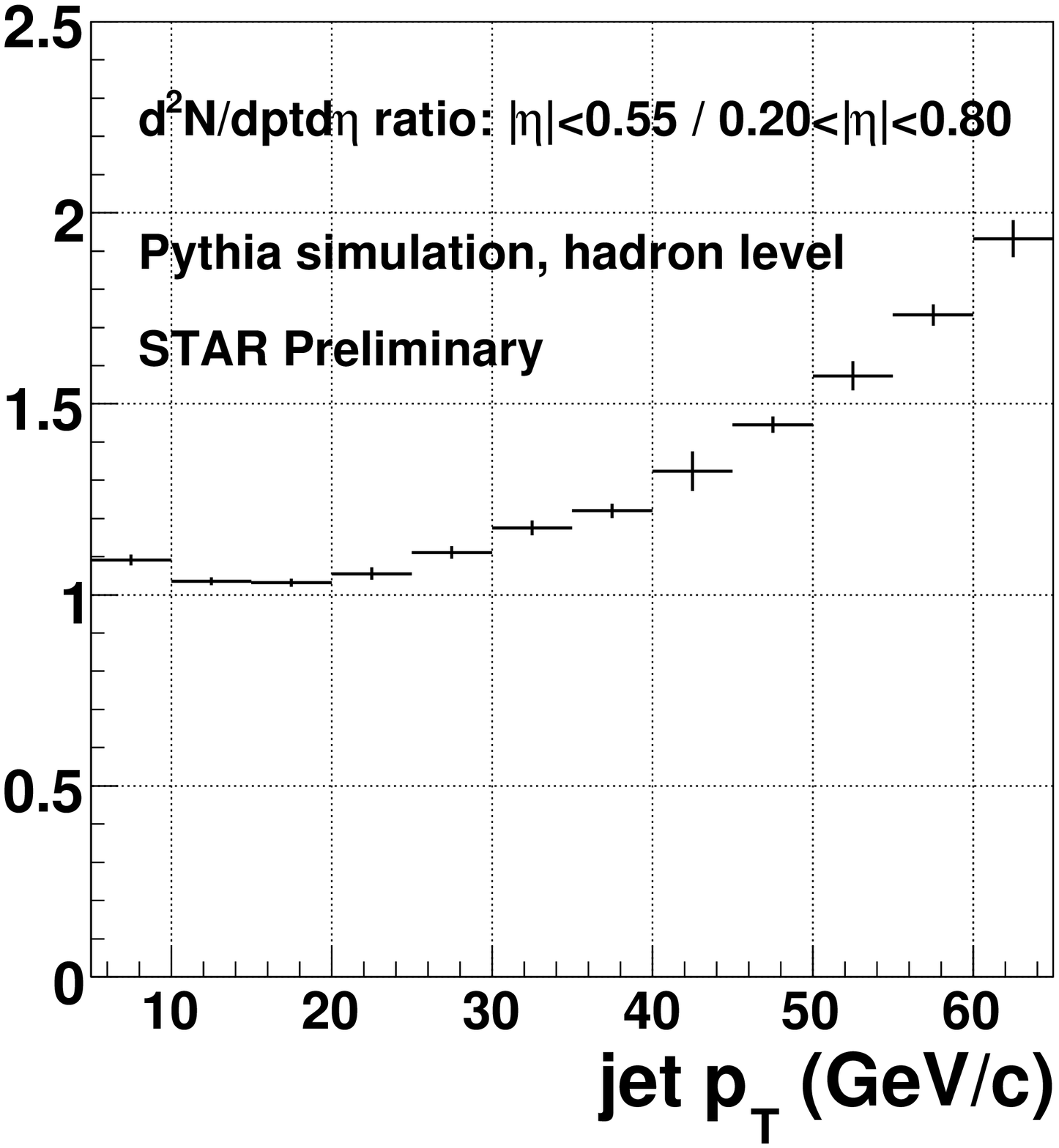}
\vspace{-0.9cm}
\caption{\label{fig:etadep}Effect of $\eta$ acceptance on jet $\pT$ spectra (Pythia).}
\end{minipage}
\end{figure}

The JES uncertainty dominates the uncertainties of d+Au measurement and is marked by the dashed lines in Fig.~\ref{fig:spectrum}. Part of it comes from the BEMC calibration uncertainty of 5\%: it's applied to the neutral component of the jet ($\approx 40\%$ for minimum-bias online trigger, $\approx 55\%$ for HT online trigger). 

An uncertainty of 10\% in TPC tracking efficiency is applied to the charged component of jets. Embedding of jets into real d+Au events at raw detector level will allow to decrease this uncertainty in the future. As JES uncertainty is expected to be largely correlated between run 8 p+p and d+Au data, we plan to measure jet $\pT$ spectrum in run 8 p+p collisions to decrease uncertainties in $R_\mathrm{dAu}$.

Caution is needed due to the use of different pseudorapidity acceptances and different jet algorithms in Fig.~\ref{fig:spectrum}. The effect of $\eta$ acceptance on jet $\pT$ spectrum is illustrated in Fig.~\ref{fig:etadep} using Pythia simulation: it is less than 10\% in the $\pT$ range covered by the present measurement. The effect of jet algorithm is mainly due to hadronization and can exceed 20\% for $\pT < 20~\gevc$~\cite{gregory}. Therefore, the same acceptance and the same algorithm have to be used in p+p and d+Au to obtain jet $R_\mathrm{dAu}$.

\section{Summary}
\label{summary}

Di-jet $\kt$ widths were measured in 200 GeV p+p and d+Au collisions: $\sigma_{k_\mathrm{T,raw}}^{p+p} = 2.8 \pm 0.1~\mathrm{(stat)}~\gevc$, $\sigma_{k_\mathrm{T,raw}}^{d+Au} = 3.0 \pm 0.1~\mathrm{(stat)}~\gevc$. No significant broadening due to Cold Nuclear Matter effects was observed.
Jet $\pT$ spectrum from minimum bias 200 GeV d+Au collisions is consistent with the scaled p+p jet spectrum within systematic uncertainties. Precise tracking efficiency determination from jet embedding in raw d+Au data and jet cross section measurement in run 8 p+p data will allow to construct jet $R_\mathrm{dAu}$. 

\section*{Acknowledgement}
\label{acknowledgement}

This work was supported in part by grants LC07048 and LA09013 of the Ministry of Education of the Czech Republic and by the grant SVV-2010-261 309.

\vspace{-0.35cm}


\end{document}